# Theoretical comparison of quantum Zeno gates and nonlinear phase gates

Hao You and J.D. Franson
*Physics Department, University of Maryland, Baltimore County, Baltimore, MD 21250*

Quantum logic operations can be implemented using nonlinear phase shifts (the Kerr effect) or the quantum Zeno effect based on strong two-photon absorption. Both approaches utilize three-level atoms, where the upper level is tuned on resonance for the Zeno gates and off-resonance for the nonlinear phase gates. The performance of nonlinear phase gates and Zeno gates are compared under conditions where the parameters of the resonant cavities and three-level atoms are the same in both cases. It is found that the expected performance is comparable for the two approaches, despite the apparent differences in the way they are implemented.

## I. INTRODUCTION

Many different approaches for implementing quantum logic gates are currently being investigated, including trapped ions [1], neutral atoms [2], polar molecules [3], superconductivity [4], solid-state spin systems [5], and optical approaches to name a few. Even within the field of quantum optics there are many different approaches, such as nonlinear phase shifts [6,7], linear optics [8-14], weak nonlinearities [15,16], photon blockade [17], continuous variables [18-20], and quantum Zeno gates [21-26]. Here we compare the expected performance of Zeno gates with logic gates based on nonlinear phase shifts (the Kerr effect). Our main goal is to understand the physical connection between these two kinds of logic gates that appear to be implemented in very different ways. This paper is not intended to provide a comparison with other cavity QED approaches, such as those based on post selection.

Nonlinear phase gates [6,7] and Zeno gates [21,22] can both be implemented using three-level atoms, as illustrated in Fig. 1. A nonlinear phase shift can be produced if photons at frequencies $\omega_1$ and $\omega_2$ are detuned from both the intermediate and upper atomic levels, as illustrated in Fig. 1a. The detuning $\delta$ from the upper level must be relatively large in order to avoid decoherence due to the decay of the upper level, as will be seen in more detail below. In contrast, Zeno gates are designed to operate with the sum of the two photon energies on resonance with the upper atomic level in order to maximize the two-photon absorption coefficient, as illustrated in Fig. 1b. This has the counter-intuitive effect of inhibiting the occurrence of two-photon absorption as a result of the Zeno effect, as described in more detail in Section III. Thus nonlinear phase gates and Zeno gates rely on changes in the real or imaginary part of the nonlinear index of refraction, respectively, and there is a strong connection between the two.

Both approaches benefit greatly from using resonant cavities to reduce the mode volume and enhance the electric fields associated with a single photon [22,27]. The resonant cavities are utilized in different ways, as will be described below. Nevertheless, we can assume that resonant cavities with the same parameters (quality factors, etc.) are used in both cases along with identical three-level atoms. This allows a direct comparison of the fidelities that should be achievable using the two approaches. It will be seen that the performances of these two kinds of gates are surprisingly similar given the differences in the way they are implemented.

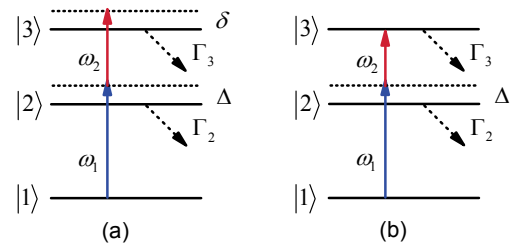

Fig. 1. (a) Nonlinear phase shift implemented using two photons detuned from the intermediate and upper levels of a three-level atom. (b) Two-photon absorption implemented with photon 1 detuned from the intermediate atomic level but with the sum of the photon energies on resonance with the upper level.

Section II describes the operation of nonlinear phase gates in more detail and calculates their expected performance characteristics. Section III performs the same calculations for quantum Zeno gates assuming the same experimental parameters. The use of atomic vapors containing a large number of atoms is investigated in Section IV, and this technique is found to have potential advantages for either kind of logic gate. The optimal performance for Zeno gates and nonlinear phase gates are compared in Section V, and a summary and conclusions are presented in Section VI.

## II. NONLINEAR PHASE GATES

Nonlinear phase shifts generated by two single photons with different frequencies were first demonstrated using a single atom inside a Fabry-Perot microcavity [7]. More recently, there has been interest in using toroidal microcavities which have a smaller mode volume and other potential advantages [28-31].

A simplified representation of a quantum phase gate using two photons with frequencies $\omega_1$ and $\omega_2$ as the qubits is illustrated schematically in Fig. 2. The incident photons are coupled into a toroidal resonator using two waveguides or tapered optical fibers. We will assume that both photons are present in the resonator at an initial time $t_0$, after which a nonlinear phase shift develops due to the Kerr effect. The interaction is allowed to continue until a nonlinear phase shift of $\pi$ has been generated, at which time the photons are assumed to be coupled out of the resonator and back into the waveguides. The feasibility of switching the photons into and out of a resonator is discussed in Appendix A. A controlled phase shift of this kind can be used to implement other quantum logic gates, such as a controlled-NOT gate, and it is sufficient for quantum computation when combined with single-photon operations.

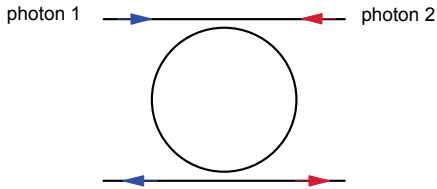

Fig. 2. Nonlinear phase gate implemented using a toroidal microcavity. Each photon is coupled into the resonator where the evanescent field couples the photons to the three-level atoms of Fig. 1a. A nonlinear phase shift of $\pi$ is produced if both photons are present, after which they are coupled out into another waveguide.

The toroidal resonator is assumed to be surrounded by an atomic vapor containing three-level atoms as illustrated in Fig. 1a. Toroidal resonators with sufficiently small minor diameters can have evanescent fields outside of the resonator that contain a substantial amount of the electromagnetic field energy [32]. This allows the photons inside the toroidal resonator to be coupled to the atoms via their evanescent fields. For simplicity, we will initially assume that the photons are coupled to a single atom. The results will then be generalized to many atoms in Section IV.

The atomic transition frequency from the ground state $|1\rangle$ to the intermediate excited state $|2\rangle$ will be denoted $\omega_{21}$, while the transition frequency from state $|2\rangle$ to the second excited state $|3\rangle$ will be denoted $\omega_{32}$. These transitions are assumed to have electric dipole moments $\mathbf{d}_1$ and $\mathbf{d}_2$, while the transition $|1\rangle \rightarrow |3\rangle$ is forbidden in the dipole approximation. The decay rates from the two excited states will be denoted by $\Gamma_2$ and $\Gamma_3$, respectively, and these include the effects of spontaneous emission as well as possible atomic collisions.

As illustrated in Fig. 1a, photon 1 is assumed to be detuned from the frequency of the first atomic transition by $\Delta$ while the sum of the two photon frequencies is detuned from the upper atomic level by $\delta$. These detunings can be written as $\Delta = \omega_1 - \omega_{21}$ and $\delta = \omega_1 + \omega_2 - \omega_{21} - \omega_{32}$. The atomic matrix elements will be denoted by $g_1 = \langle -\mathbf{d}_1 \cdot \hat{\mathbf{E}}_1(\vec{r}) \rangle$ and $g_2 = \langle -\mathbf{d}_2 \cdot \hat{\mathbf{E}}_2(\vec{r}) \rangle$, which correspond to transitions between states $|1\rangle \rightarrow |2\rangle$ and $|2\rangle \rightarrow |3\rangle$, respectively. Here $\hat{\mathbf{E}}_1$ and $\hat{\mathbf{E}}_2$ are the electric field operators associated with the photons at frequency $\omega_1$ and $\omega_2$, respectively, at the location of the atom. They can be evaluated using the classical field modes as described in Ref. [32]. The matrix elements are assumed to be averaged over all possible locations of the atoms and orientations of their dipoles.

In the absence of any losses, the shift in the ground-state energy of the atom could be calculated in a straightforward way using perturbation theory. We will assume that the system is initially in a pure state, but the presence of the decay rates $\Gamma_2$ and $\Gamma_3$ will then produce a mixed state at the output of the device. We previously showed [33] that the density matrix can be factored under these conditions, which simplifies the analysis.

Here we will use an equivalent perturbation theory approach in which the detunings include factors of $i\Gamma_2/2$ and $i\Gamma_3/2$. For example, it can be shown [34-36] that the shift in the energy of the ground state is effectively given by

$$E^{(4)} = \frac{|g_1|^2 |g_2|^2}{\hbar^3 \left(\Delta + \frac{i}{2}\Gamma_2\right)^2 \left(\delta + \frac{i}{2}\Gamma_3\right)} \qquad (1)$$

The real part of this expression gives the actual shift in the energy while its imaginary part reflects the reduced lifetime of the ground state due to the decay rates of the upper levels to which it is coupled. Similar expressions exist for the probability amplitudes of the excited states. It can be shown using the method of resolvents [35] that this is equivalent to the density-matrix calculations of Ref. [33] in the perturbative limit of weak couplings.

Since photon loss is the dominant decoherence mechanism, we will characterize the performance of the devices by the probability $P_S$ that no photons were absorbed during the operation of the gate, which we will refer to as the success probability. This is equivalent to the fidelity of the output states if all other error sources are negligible. The photon loss is largest when two photons are present, and for simplicity we will limit the comparison to that case.

In a perturbative approach, the coupling is assumed to be sufficiently weak that there is no significant change in the probability $\rho_{11}$ that the atom will remains in state $|1\rangle$ with two photons present at frequency $\omega_1$ and $\omega_2$, so that $\rho_{11} = 1$. The probability amplitude that the atom will absorb photon 1 and make a transition to state $|2\rangle$ corresponds to a first-order correction $|\psi_1^{(1)}\rangle$ to the atomic state given by:

$$|\psi_1^{(1)}\rangle = \frac{g_1}{\hbar(\Delta + i\Gamma_2/2)}|2\rangle. \quad (2)$$

The corresponding probability for this state is

$$\rho_{22} = \langle\psi_1^{(1)}|\psi_1^{(1)}\rangle = \frac{4|g_1|^2}{\hbar^2(4\Delta^2 + \Gamma_2^2)} \quad (3)$$

Similarly, the perturbation may cause the atom to absorb both photons and make a transition to state $|3\rangle$, which leads to a second order perturbed state $|\psi_1^{(2)}\rangle$ given by

$$|\psi_1^{(2)}\rangle = \frac{g_1 g_2}{\hbar^2(\Delta + i\Gamma_2/2)(\delta + i\Gamma_3/2)}|3\rangle. \quad (4)$$

The corresponding probability for this state is

$$\rho_{33} = \langle\psi_1^{(2)}|\psi_1^{(2)}\rangle = \frac{16|g_1 g_2|^2}{\hbar^4(4\Delta^2 + \Gamma_2^2)(4\delta^2 + \Gamma_3^2)}. \quad (5)$$

The shift in the energy of the system as a result of these virtual transitions is given by the real part of Eq. (1):

$$\text{Re}\, E^{(4)} = \frac{16|g_1|^2|g_2|^2}{\hbar^3(4\Delta^2 + \Gamma_2^2)^2(4\delta^2 + \Gamma_3^2)} \times (4\delta\Delta^2 - 2\Gamma_2\Gamma_3\Delta - \Gamma_2^2\delta). \quad (6)$$

This energy shift only occurs if both photons are present in the cavity and it can be used to realize a $\pi$ phase shift in a controlled phase gate, which is also referred to as a controlled-Z (CZ) gate. In order to achieve this, the photons must interact inside the resonator for a time interval $t_p$ given by

$$t_p = \frac{\pi}{|\text{Re}\, E^{(4)}/\hbar|} = \frac{\pi\hbar^4(4\Delta^2 + \Gamma_2^2)^2(4\delta^2 + \Gamma_3^2)}{16|g_1|^2|g_2|^2|4\delta\Delta^2 - 2\Gamma_2\Gamma_3\Delta - \Gamma_2^2\delta|}. \quad (7)$$

To evaluate the performance of the logic gate, we need to determine the probability that one or more photons will be lost during the time interval $t_p$ required to perform a nonlinear $\pi$ phase shift. Loss can occur in two different ways, either by direct photon absorption in the cavity or as the result of a decay of one of the excited atomic states.

As usual [37], we will define the rate at which a photon is absorbed or lost by some other mechanism in the cavity by a decay rate $\kappa$, whose value can be determined from the quality factor Q of the cavity. This type of loss can occur in both the original two-photon state or in a virtual state with only a single photon present. The probability of a cavity photon loss in the two-photon state during the time interval $t_p$ is given by

$$\rho_{11} \times 2\kappa t_p = \frac{2\kappa\pi\hbar^4(4\Delta^2 + \Gamma_2^2)^2(4\delta^2 + \Gamma_3^2)}{16|g_1|^2|g_2|^2|4\delta\Delta^2 - 2\Gamma_2\Gamma_3\Delta - \Gamma_2^2\delta|}. \quad (8)$$

The probability of a cavity photon loss in the single-photon virtual state is

$$\rho_{22}\kappa t_p = \frac{\kappa\pi\hbar^2(4\Delta^2 + \Gamma_2^2)(4\delta^2 + \Gamma_3^2)}{4|g_2|^2|4\delta\Delta^2 - 2\Gamma_2\Gamma_3\Delta - \Gamma_2^2\delta|}. \quad (9)$$

Decay of the atomic state can occur in both the intermediate and upper levels. During the time interval $t_p$, the probability that the intermediate atomic level will decay is given by

$$\rho_{22}\Gamma_2 t_p = \frac{\Gamma_2\pi\hbar^2(4\Delta^2 + \Gamma_2^2)(4\delta^2 + \Gamma_3^2)}{4|g_2|^2|4\delta\Delta^2 - 2\Gamma_2\Gamma_3\Delta - \Gamma_2^2\delta|}. \quad (10)$$

The probability of a decay of the upper atomic level is

$$\rho_{33}\Gamma_3 t_p = \frac{\pi\Gamma_3(4\Delta^2 + \Gamma_2^2)}{|4\delta\Delta^2 - 2\Gamma_2\Gamma_3\Delta - \Gamma_2^2\delta|} \quad (11)$$

Thus the total probability for a failure event in which a $\pi$ phase shift was not properly performed is $P_F = \rho_{22}\Gamma_2 t_p + \rho_{33}\Gamma_3 t_p + 2\kappa t_p + \rho_{22}\kappa t_p$. This corresponds to a probability of success given by $P_S = 1 - P_F$, or

$$P_S = 1 - \frac{\pi \Gamma_3 \left(4\Delta^2 + \Gamma_2^2\right)}{\left|4\delta\Delta^2 - 2\Gamma_2\Gamma_3\Delta - \Gamma_2^2\delta\right|}$$

$$- \frac{2\kappa\pi\hbar^4 \left(4\Delta^2 + \Gamma_2^2\right)^2 \left(4\delta^2 + \Gamma_3^2\right)}{16|g_1 g_2|^2 \left|4\delta\Delta^2 - 2\Gamma_2\Gamma_3\Delta - \Gamma_2^2\delta\right|} \quad (12)$$

$$- \frac{\pi\hbar^2 (\kappa + \Gamma_2)\left(4\Delta^2 + \Gamma_2^2\right)\left(4\delta^2 + \Gamma_3^2\right)}{4|g_2|^2 \left|4\delta\Delta^2 - 2\Gamma_2\Gamma_3\Delta - \Gamma_2^2\delta\right|}.$$

In order to simplify the calculations and various trade-offs in the design of the gates, we will assume that $\Gamma_2 = \Gamma_3 = \Gamma$ and that $g_1 = g_2 = g$. These assumptions are only for convenience and not necessary. It will also be convenient to introduce dimensionless parameters defined by

$$\begin{aligned}
\delta_r &= \delta/\Gamma \\
\Delta_r &= \Delta/\Gamma \\
\Gamma_r &= \Gamma/\kappa \\
\Omega &= g/(\hbar\Gamma).
\end{aligned} \quad (13)$$

Here $\delta_r = \delta/\Gamma$ and $\Delta_r = \Delta/\Gamma$ are the relative upper and intermediate level detunings as compared to the atomic decay rate while $\Gamma_r = \Gamma/\kappa$ is the ratio of the atomic decay to the cavity mode loss. $\Omega = g/(\hbar\Gamma)$ is the relative Rabi frequency as compared to the atomic decay rate and it is proportional to the inverse square root of the usual "saturation" photon number [37].

In terms of these dimensionless parameters, Eq.(12) now becomes

$$P_S = 1 - \frac{\pi\left(4\Delta_r^2 + 1\right)}{\left|4\delta_r\Delta_r^2 - 2\Delta_r - \delta_r\right|}$$

$$- \frac{\pi\left(4\Delta_r^2 + 1\right)^2 \left(4\delta_r^2 + 1\right)}{8\Gamma_r|\Omega|^4 \left|4\delta_r\Delta_r^2 - 2\Delta_r - \delta_r\right|} \quad (14)$$

$$- \frac{\pi(1+\Gamma_r)\left(4\Delta_r^2 + 1\right)\left(4\delta_r^2 + 1\right)}{4\times\Gamma_r|\Omega|^2 \left|4\delta_r\Delta_r^2 - 2\Delta_r - \delta_r\right|}.$$

The above results can be used to determine the expected performance of a nonlinear phase gate for a given set of parameters. The detunings can be varied in order to optimize the performance of the device for a given set of hardware parameters, as will be described in Section V.

### III. QUANTUM ZENO GATES

A controlled phase shift of $\pi$ can also be produced using two-photon absorption and the quantum Zeno effect [21,22]. As illustrated in Fig. 3, single photons present in the waveguides are assumed to be coupled into two toroidal resonators at an initial time $t_0$. (The ability to switch the photons into and out of a resonator is discussed in the Appendix.) The two resonators are coupled to each other by their evanescent fields, so that if only one photon is present in one of the resonators it will be completely transferred to the other resonator after a time interval $t_s$. This transfer process is analogous to the Rabi oscillations between two atomic states and it produces a linear phase shift of $\pi/2$ for states $|01\rangle$ or $|10\rangle$ where there is a single photon initially present in only one of the resonators; the same $\pi/2$ phase shift also occurs for an atomic Rabi oscillation [21].

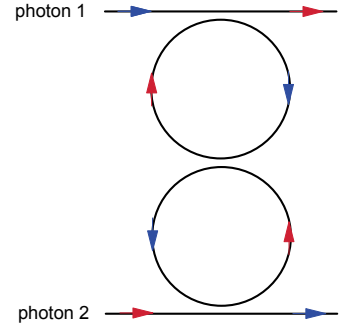

Fig.3. Quantum Zeno gate implemented using two toroidal microcavities [22]. If only one photon is present in one of the resonators, it will be coupled into the other resonator producing a phase shift of $\pi/2$ in the process. If one photon is present in each of the resonators, strong two-photon absorption will inhibit the coupling of the photons between the resonators and eliminate the phase shift of $\pi/2$ for each photon that would otherwise occur. (A dual-rail encoding is assumed with the other two paths not shown.)

As described in more detail in Ref. [21], the basic idea is to use two-photon absorption to inhibit the growth of any probability amplitude for two photons to be in the same resonator. Consider an initial state $|11\rangle$ containing one photon initially present in both resonators. In that case, neither of the photons can couple into the other resonator in the limit of strong two-photon absorption, since that would give a state with two photons in the same resonator which is suppressed. This eliminates the two phase shifts of $\pi/2$ that would otherwise occur in the absence of any interaction between the photons, which gives a net nonlinear phase shift of $\pi$ as compared to the input states $|01\rangle$ or $|10\rangle$. The fact that a single photon will be transferred to the other resonator can be compensated by simply swapping the output paths.

The nonlinear phase shift produced by a Zeno gate can be combined with two Hadamard operations (beam splitters) to implement a CNOT gate. This is effectively an interferometer where the reduction in probability amplitudes due to photon loss can unbalance the interferometer. Photon loss itself can be corrected using a relatively simple error correction code [25] but the errors due to unbalanced interferometers would require the use of a more general error correction code. Leung et al. [23,24] have shown that these difficulties can be avoided by deliberately introducing loss into the other paths of the interferometer. Here we only calculate the decreased fidelity due to photon loss in the Zeno gate itself. This is sufficient for comparison with nonlinear phase gates, since similar difficulties with unbalanced interferometers will occur there as well.

The coupling between the two resonators can be described by the Hamiltonian

$$\hat{H}_c = \hbar\omega\left(\hat{a}_A^+\hat{a}_A + \hat{a}_B^+\hat{a}_B\right) + \varepsilon\left(\hat{a}_A^+\hat{a}_B + \hat{a}_B^+\hat{a}_A\right) \quad (15)$$

Here the operators $\hat{a}_A^\dagger$ and $\hat{a}_B^\dagger$ create a photon in the corresponding resonator while $\varepsilon$ reflects the coupling strength between the two microcavities due to the evanescent field. The magnitude of $\varepsilon$ can be controlled experimentally by adjusting the separation between the two resonators. The photons inside the resonators are once again coupled to a three-level atom via their evanescent fields. The relevant atomic states are the same as before but the sum of the frequencies of the two photons is now assumed to be on resonance, as illustrated in Fig. 1 (b).

The time required to transfer (swap) a single photon from one resonator to the other is given by $t_s = \pi\hbar/2\varepsilon$. During this time there can be decoherence due to single photon losses in the cavity as well as decay of the virtually excited atomic states. In addition, some amount of two-photon absorption will occur if the two-photon absorption coefficient is not sufficiently large.

We calculated the time evolution of the density matrix $\hat{\rho}$ using the techniques of Ref. [33]. When two photons are present in the system, the Hilbert space includes the states $|1,1\rangle$, $|2,0\rangle$, and $|0,2\rangle$, which correspond to one photon in each resonator, two photons in the upper resonator, and two photons in the lower resonator, respectively. This gave an effective transition matrix for a quantum Zeno gate in the computational (reduced) basis $(|00\rangle, |01\rangle, |10\rangle, |11\rangle)$ that is given by

$$\hat{U} = \begin{pmatrix} 1 & 0 & 0 & 0 \\ 0 & 0 & -ie^{-r_1 t_s/2} & 0 \\ 0 & -ie^{-r_1 t_s/2} & 0 & 0 \\ 0 & 0 & 0 & \alpha(t_s)e^{-(R_1+R_2/4)t_s} \end{pmatrix} \quad (16)$$

Here $\alpha(t_s) = \cosh(\Omega_0 t_s) + R_2 \sinh(\Omega_0 t_s)/4\Omega_0$, the rate of linear loss in the state $|11\rangle$ is $R_1 = \kappa + \rho_{22}(\kappa + \Gamma_2)$, and the rate of linear loss in the states $(|01\rangle, |10\rangle)$ is $r_1 = \kappa + \rho_{22}\Gamma_2$. As before, the cavity photon decay rate is $\kappa$ and the decay rate of state $|2\rangle$ is $\Gamma_2$. $R_2$ is the two-photon absorption rate in the states $|2,0\rangle$ and $|0,2\rangle$, which is given by

$$R_2 = \frac{16|g_1 g_2|^2}{\hbar^4(4\Delta^2 + \Gamma_2^2)\Gamma_3}. \quad (17)$$

Here we define

$$\Omega_0 = \frac{\sqrt{R_2^2 \hbar^2 - 64\varepsilon^2}}{4\hbar}. \quad (18)$$

Allowing a sufficient time $t_s$ for a single photon to be completely transferred (swapped) from one resonator to the other gives

$$R_1 t_s = \left[\kappa + \rho_{22}(\kappa + \Gamma_2)\right]\frac{\pi\hbar}{2\varepsilon}$$
$$= \left[\kappa + \frac{4|g_1|^2}{\hbar^2(4\Delta^2 + \Gamma_2^2)}(\kappa + \Gamma_2)\right]\frac{\pi\hbar}{2\varepsilon}, \quad (19)$$

$$R_2 t_s = \frac{8\pi|g_1 g_2|^2}{\hbar^3(4\Delta^2 + \Gamma_2^2)\Gamma_3\varepsilon}, \quad (20)$$

and

$$\Omega_0 t_s = \frac{\sqrt{R_2^2\hbar^2 - 64\varepsilon^2}}{4\hbar}t_s = \sqrt{\frac{R_2^2 t_s^2}{16} - \pi^2}. \quad (21)$$

In addition to the dimensionless parameters already defined in Eq. (13), we also introduce a new variable $\varepsilon_\kappa$ that describes the coupling strength between two resonators as compared $\kappa$:

$$\varepsilon_\kappa = \frac{\varepsilon/\hbar}{\kappa} \quad (22)$$

In terms of these dimensionless parameters

$$R_1 t_s = \frac{\pi}{2\varepsilon_\kappa}\left[1 + \frac{4(1+\Gamma_r)|\Omega|^2}{4\Delta_r^2 + 1}\right] \quad (23)$$

$$R_2 t_s = \frac{8\pi \Gamma_r \Omega^4}{\varepsilon_\kappa \left(4\Delta_r^2 + 1\right)} \quad (24)$$

and

$$\Omega_0 t_s = \pi \sqrt{\frac{4\Gamma_r^2 \Omega^8}{\varepsilon_\kappa^2 \left(4\Delta_r^2 + 1\right)^2} - 1}. \quad (25)$$

In addition

$$\frac{R_2}{4\Omega_0} = \frac{R_2 t_s}{4\Omega_0 t_s} = \frac{R_2 t_s}{4\sqrt{(R_2 t_s)^2/16 - \pi^2}}$$

$$= \frac{R_2 t_s}{\sqrt{(R_2 t_s)^2 - 16\pi^2}}. \quad (26)$$

Combining these results gives the fidelity $\langle 1,1|\hat{\rho}|1,1\rangle$ corresponding to the desired output state $|11\rangle$ with the nonlinear phase shift applied. This is equivalent to a success probability of

$$P_s = e^{-2(R_1 + R_2/4)t_s} [\cosh(\Omega_0 t_s) + \frac{R_2}{4\Omega_0}\sinh(\Omega_0 t_s)]^2. \quad (27)$$

The coupling constant $\varepsilon$ can be varied in order to optimize the performance of a Zeno logic gate, as will be described in more detail in Section V. The optimized performance of nonlinear phase gates and Zeno gates will then be compared in Section VI.

## IV. USE OF ATOMIC VAPOR

For simplicity, the calculations in the previous two sections assumed that each resonator was coupled to a single atom, which can be achieved experimentally by trapping an atom in the evanescent field of the resonators. In this section, we generalize the results to the case in which the resonators are coupled to $N$ atoms in an atomic vapor. This is necessary for the operation of quantum Zeno gates, and it will be seen that it has potential advantages for nonlinear phase gates as well. In both approaches, using a large number of atoms reduces the time interval required to achieve the desired logic operation, which tends to reduce losses associated with cavity decay modes.

First consider the case of a nonlinear phase gate coupled to $N$ atoms in the evanescent field. We average over the locations of the atoms and replace the coupling constants $g_1$ and $g_2$ with their effective values. The nonlinear shift in the energy of the system now becomes

$$N \operatorname{Re} E^{(4)} = \frac{16 N |g_1|^2 |g_2|^2}{\hbar^3 \left(4\Delta^2 + \Gamma_2^2\right)^2 \left(4\delta^2 + \Gamma_3^2\right)} \quad (28)$$

$$\times \left(4\delta\Delta^2 - 2\Gamma_2\Gamma_3\Delta - \Gamma_2^2\delta\right)$$

To realize a $\pi$ phase shift now requires a time interval given by

$$t_{Np} = \frac{\pi}{N\left|\operatorname{Re} E^{(4)}\right|/\hbar}$$

$$= \frac{\pi \hbar^4 \left(4\Delta^2 + \Gamma_2^2\right)^2 \left(4\delta^2 + \Gamma_3^2\right)}{16 N |g_1|^2 |g_2|^2 \left|4\delta\Delta^2 - 2\Gamma_2\Gamma_3\Delta - \Gamma_2^2\delta\right|}. \quad (29)$$

The losses are due once again to the decay of the photons inside the cavity and the decay of the virtually-excited atomic states. The cavity decay rate rate $\kappa$ is unaffected by the number of atoms, so that the photon loss probability in the two-photon state is

$$\rho_{11} \times 2\kappa t_{Np} = \frac{2\kappa\pi\hbar^4 \left(4\Delta^2 + \Gamma_2^2\right)^2 \left(4\delta^2 + \Gamma_3^2\right)}{16 N |g_1|^2 |g_2|^2 \left|4\delta\Delta^2 - 2\Gamma_2\Gamma_3\Delta - \Gamma_2^2\delta\right|}. \quad (30)$$

The photon loss probability in the single-photon state with one excited atom is now

$$\rho_{22} \kappa t_{Np} = \frac{\kappa\pi\hbar^2 \left(4\Delta^2 + \Gamma_2^2\right)\left(4\delta^2 + \Gamma_3^2\right)}{4N|g_2|^2 \left|4\delta\Delta^2 - 2\Gamma_2\Gamma_3\Delta - \Gamma_2^2\delta\right|}. \quad (31)$$

Note that both of these loss rates are reduced by a factor of $1/N$ due to the reduced gate operation time.

The total probability of an excited atom is increased by a factor of $N$ in the perturbative limit, while the corresponding losses due to atomic decay are reduced by a factor of $1/N$ due to the reduced operation time. As a result, the losses due to decay in the intermediate atomic state becomes

$$N\rho_{22}\Gamma_2 t_{Np} = \frac{\Gamma_2 \pi\hbar^2 \left(4\Delta^2 + \Gamma_2^2\right)\left(4\delta^2 + \Gamma_3^2\right)}{4|g_2|^2 \left|4\delta\Delta^2 - 2\Gamma_2\Gamma_3\Delta - \Gamma_2^2\delta\right|} \quad (32)$$

while the losses in the upper atomic state are given by

$$N\rho_{33}\Gamma_3 t_{Np} = \frac{\pi\Gamma_3 \left(4\Delta^2 + \Gamma_2^2\right)}{\left|4\delta\Delta^2 - 2\Gamma_2\Gamma_3\Delta - \Gamma_2^2\delta\right|}. \quad (33)$$

Thus the total failure probability for a nonlinear phase gate is given by

$$P_f^{(\pi)} = N\rho_{22}\Gamma_2 t_{Np} + N\rho_{33}\Gamma_3 t_{Np} + 2\kappa t_{Np} + \rho_{22}\kappa t_{Np}.$$

The corresponding probability of success becomes

$$P_S = 1 - \frac{\pi\Gamma_3\left(4\Delta^2 + \Gamma_2^2\right)}{\left|4\delta\Delta^2 - 2\Gamma_2\Gamma_3\Delta - \Gamma_2^2\delta\right|}$$
$$- \frac{2(\kappa/N)\pi\hbar^4\left(4\Delta^2 + \Gamma_2^2\right)^2\left(4\delta^2 + \Gamma_3^2\right)}{16|g_1 g_2|^2\left|4\delta\Delta^2 - 2\Gamma_2\Gamma_3\Delta - \Gamma_2^2\delta\right|} \quad (34)$$
$$- \frac{\pi\hbar^2(\kappa/N + \Gamma_2)\left(4\Delta^2 + \Gamma_2^2\right)\left(4\delta^2 + \Gamma_3^2\right)}{4|g_2|^2\left|4\delta\Delta^2 - 2\Gamma_2\Gamma_3\Delta - \Gamma_2^2\delta\right|}.$$

This can be rewritten using the dimensionless parameters of Eq. (13) as

$$P_S = 1 - \frac{\pi\left(4\Delta_r^2 + 1\right)}{\left|4\delta_r\Delta_r^2 - 2\Delta_r - \delta_r\right|}$$
$$- \frac{\pi\left(4\Delta_r^2 + 1\right)^2\left(4\delta_r^2 + 1\right)}{8N\Gamma_r|\Omega|^4\left|4\delta_r\Delta_r^2 - 2\Delta_r - \delta_r\right|} \quad (35)$$
$$- \frac{\pi(1 + N\Gamma_r)\left(4\Delta_r^2 + 1\right)\left(4\delta_r^2 + 1\right)}{4N\Gamma_r|\Omega|^2\left|4\delta_r\Delta_r^2 - 2\Delta_r - \delta_r\right|}.$$

Coupling the photons to a large number of atoms has similar effects on the operation of a quantum Zeno gate. Roughly speaking, the coupling between the two resonators of Fig. 3 has to be slow compared to the rate of two-photon absorption in order to minimize the error rate. Increasing the number of atoms allows the time $t_s$ required for the operation of the gate to be reduced while maintaining the same error rate due to two-photon absorption, although the relation $t_s = \pi\hbar/2\varepsilon$ still holds.

The linear loss rate in the state $|11\rangle$ is now given by

$$R_{N1} = \kappa + \rho_{22}(\kappa + N\Gamma_2) \quad (36)$$

while the two-photon absorption rate in states $|2,0\rangle$ and $|0,2\rangle$ becomes

$$R_{N2} = \frac{16N|g_1 g_2|^2}{\hbar^4\left(4\Delta^2 + \Gamma_2^2\right)\Gamma_3}. \quad (37)$$

Introducing the dimensionless parameters of Eqs. (13) and Eq. (22) gives

$$R_{N1}t_s = \frac{\pi}{2\varepsilon_\kappa}\left[1 + \frac{4(1 + N\Gamma_r)|\Omega|^2}{4\Delta_r^2 + 1}\right], \quad (38)$$

$$R_{N2}t_s = \frac{8\pi N\Gamma_r \Omega^4}{\varepsilon_\kappa\left(4\Delta_r^2 + 1\right)}, \quad (39)$$

and

$$\Omega_N t_s = \pi\sqrt{\frac{4N^2\Gamma_r^2\Omega^8}{\varepsilon_\kappa^2\left(4\Delta_r^2 + 1\right)^2} - 1}. \quad (40)$$

This can also be written as

$$\frac{R_{N2}}{4\Omega_N} = \frac{R_{N2}t_s}{4\Omega_N t_s} = \frac{R_{N2}t_s}{4\sqrt{(R_{N2}t_s)^2/16 - \pi^2}}$$
$$= \frac{R_{N2}t_s}{\sqrt{(R_{N2}t_s)^2 - 16\pi^2}}. \quad (41)$$

For two photons initially in the state $|11\rangle$, this corresponds to a success probability given by

$$P_s = e^{-2(R_{N1} + R_{N2}/4)t_s}[\cosh(\Omega_N t_s)$$
$$+ \frac{R_{N2}}{4\Omega_N}\sinh(\Omega_N t_s)]^2. \quad (42)$$

## V. PERFORMANCE COMPARISONS

The results of the previous sections give the expected performance of Zeno gates and nonlinear phase gates as functions of the various experimental parameters. Before we can compare their performance, however, we need to optimize the performance of each type of logic gate by choosing the optimal value of any parameters that could reasonably be controlled experimentally. It should be noted that we are comparing the performance of these two kinds of gates in order to understand the physical connection between the two, and that this paper is not intended to be a review of the potential performance of other cavity QED approaches.

We assume that the best available resonators will be used in both cases, so that the value of $\kappa$ cannot be chosen arbitrarily. In addition, we assume that the same species of atoms are used in both approaches, so that $g_1$, $g_2$, $\omega_{12}$, $\omega_{23}$, $\Gamma_2$, and $\Gamma_3$ all have fixed values. This leaves the detunings $\delta$ and $\Delta$ as well as the resonator coupling $\varepsilon$ as parameters that can be controlled experimentally in order to optimize the performance of both logic gates as appropriate.

For simplicity, we will primarily consider a single baseline set of cavity parameters, which we arbitrarily take to be $\Gamma_r = 0.1$ and $\Omega = 50$. This corresponds to the strong coupling regime [37,38], which enhances the performance of both types of logic gates in essentially the same way. Similar comparison results should be expected for other choices of the cavity parameters, as will be discussed later in this section. We will first compare the performances of the two devices for the case of a single trapped atom and then consider the effects of using $N$ atoms.

The optimal performance of nonlinear phase gates depends on the dimensionless detuning parameters $\delta_r$ and $\Delta_r$. Fig. 4a shows a plot of the success probability $P_s$ as a function of $\delta_r$ with the intermediate state detuning fixed at $\Delta_r = 6$ or $\Delta_r = 20$. Reducing the value of the detuning $\delta_r$ initially increases the coupling into the upper atomic state responsible for the nonlinear phase shift, which reduces the required operation time and the total loss. But reducing the value of $\delta_r$ beyond an optimal value eventually increases the population of the upper atomic level to the point that its losses dominate and the total success rate begins to decrease.

We can see from Fig. 4a that the maximum probability of success also depends on the value of the intermediate detuning $\Delta_r$ and both detunings must be varied in order to optimize the performance. Fig. 4b shows a two-dimensional plot of the probability of success of a nonlinear phase gate as function of the detuning parameters $\delta_r$ and $\Delta_r$. The maximum success probability of approximately 0.57 is achieved for $\delta_r = 14.9$ and $\Delta_r = 6.4$ for this choice of cavity parameters. Once again, these results correspond to the presence of a single atom.

We now consider the optimal performance of quantum Zeno gates assuming the same values as before for all of the fixed parameters. But here we choose $\delta = 0$ in order to maximize the two-photon absorption coefficient, while the resonator coupling parameter $\varepsilon_\kappa$ can be varied in order to optimize the performance.

The dashed line in Fig. 5 shows the success probability for a quantum Zeno gate as a function of the coupling parameter $\varepsilon_\kappa$ with the intermediate detuning parameter $\Delta_r$ fixed at a value 6.4, which corresponds to its optimal value for the nonlinear phase gates. In the limit of large coupling between the two resonators, the two-photon absorption coefficient is not strong enough to produce a sufficiently large Zeno effect during the short time interval required to swap a photon. On the other hand, if the coupling is too weak, the logic operation will take more time and other loss mechanisms will reduce the success rate. An optimal success rate of 0.55 is achieved for a value of $\varepsilon_\kappa$ approximately equal to 725. We note that this is very close to the optimal success rate of 0.57 for a nonlinear phase gate under the same conditions.

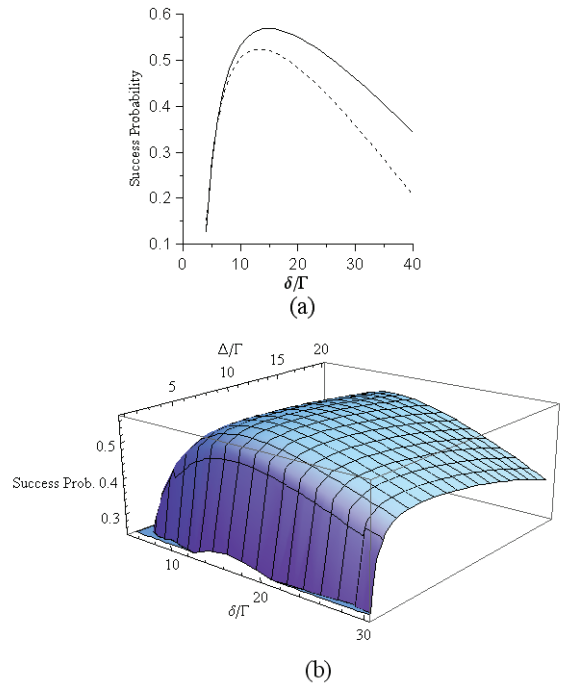

Fig.4. (a) Success probability for a nonlinear phase gate as a function of the upper level detuning parameter $\delta_r$. Here the detuning parameter for the intermediate state was fixed at $\Delta_r = 6$ (solid line) or $\Delta_r = 20$ (dashed line). (b) Two-dimensional plot of the success probability for a nonlinear phase gate as a function of both detuning parameters.

The solid line in Fig. 5 shows a plot of the success probability for a quantum Zeno gate as a function of the resonator coupling parameter $\varepsilon_\kappa$, where now the value of $\Delta_r$ was chosen to optimize the results for each value of $\varepsilon_\kappa$. It can be seen that the performance of a Zeno gate is relatively insensitive to the value of $\varepsilon_\kappa$ provided that $\Delta_r$ is adjusted accordingly. The optimal value of the success probability saturates at a value of approximately 0.6. Comparing this to Fig. 4b, we see that a Zeno gate can achieve approximately the same performance by varying the resonator coupling and intermediate detuning that a nonlinear phase gate can achieve by varying the detuning of the upper level.

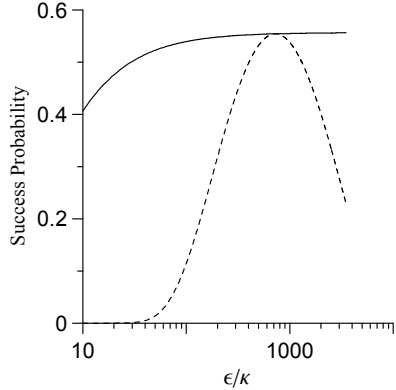

Fig. 5. Success probability for a quantum Zeno gate as a function of the dimensionless coupling parameter $\varepsilon_\kappa = \varepsilon/\kappa$ that determines the strength of the coupling between the two resonators of Fig. 3. The dashed line corresponds to a fixed value of the intermediate state detuning of $\Delta_r = 6.4$, while the solid line corresponds to $\Delta_r$ set to its optimal value for each value of $\varepsilon_\kappa$.

Although we have assumed that the cavity parameters are fixed for comparison purposes, it is interesting nevertheless to consider the effects of the normalized atomic Rabi frequency on the performance of the logic gates. This corresponds to varying the strength of the photon-atom coupling which could be achieved by reducing the mode volume, for example. We first consider the effects of the Rabi frequency on nonlinear phase gates. Fig. 6a shows the dependence of the success probability of a nonlinear phase gate on the normalized Rabi frequency $\Omega$. Here the red solid curve corresponds to the situation where the other parameters were fixed at $\delta_r = 10$, $\Delta_r = 6.4$ and $\Gamma_r = 0.1$. The blue dashed curve corresponds to the situation where the optimal values of $\delta_r$ and $\Delta_r$ were computed at each point with $\Gamma_r = 0.1$. It can be seen that increasing the strength of the photon-atom coupling would increase the success probability, as would be expected.

Similar results for quantum Zeno gates are shown in Fig. 6b, where the red solid curve corresponds to fixed parameters of $\varepsilon_\kappa = 800$, $\Delta_r = 6.4$ and $\Gamma_r = 0.1$. The blue dashed curve corresponds to optimized values of $\varepsilon_\kappa$ and $\Delta_r$ for each value of $\Omega$. It can be seen that increasing the Rabi frequency can sometimes make the performance worse if the values of $\varepsilon_\kappa$ and $\Delta_r$ are not optimized.

Fig. 6c directly compares the optimized success probability of Zeno gates to that of nonlinear phase gates as a function of the normalized Rabi frequencies. The red solid curve corresponds to a nonlinear phase gate with the optimal values of $\delta_r$ and $\Delta_r$ for each value of $\Omega$. The blue dashed curve corresponds to the success probability for a Zeno gate using the optimal values of $\varepsilon_\kappa$ and $\Delta_r$ for each value of $\Omega$. It can be seen that the performance of the two kinds of gates is remarkably similar.

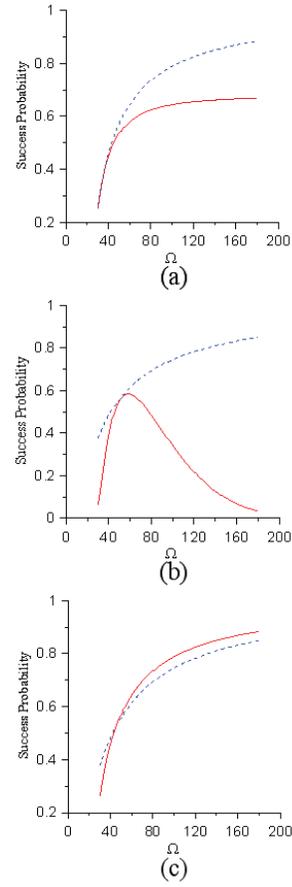

Fig. 6. Success probability of quantum logic gates as a function of the normalized atomic Rabi frequency $\Omega$ with $\Gamma_r = 0.1$. (a) Success probability for a nonlinear phase gate for fixed values of the other parameters (red solid line) and for optimized values of the other parameters (blue dashed line). (b) Success probability for a quantum Zeno gate for fixed values of the other parameters (red solid line) and for optimized values of the other parameters (blue dashed line). (c) Direct comparison of the optimized success probability for a nonlinear phase gate (red solid line) and a Zeno gate (blue dashed line).

It is also interesting to consider the effects of the cavity loss rate on the relative performance of the logic gates. Recent progress in cavity fabrication has reduced the cavity mode loss and thus increased the value of $\Gamma_r = \Gamma/\kappa$ to more typical values of $\Gamma_r = 1.8$ [31]. Fig. 7a shows the optimized success probability of nonlinear phase gates as a function of the normalized Rabi frequencies for the case of $\Gamma_r = 1.8$ and $\Gamma_r = 0.1$. Fig. 7b shows the corresponding results for quantum Zeno gates. It can be seen that the performance of both kinds of gates is improved when the cavity mode loss is decreased (larger values of $\Gamma_r$).

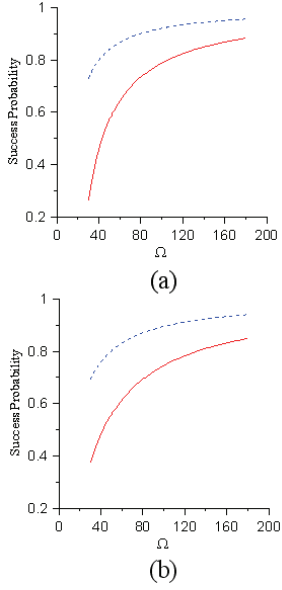

Fig. 7. Success probability of quantum logic gates as a function of the normalized atomic Rabi frequency $\Omega$ for two different values of $\Gamma_r$. (a) Success probability for a nonlinear phase gate for optimized values of the other parameters with $\Gamma_r = 0.1$ (red solid line) and with $\Gamma_r = 1.8$ (blue dashed line). (b) Success probability for a quantum Zeno gate for optimized values of the other parameters with $\Gamma_r = 0.1$ (red solid line) and with $\Gamma_r = 1.8$ (blue dashed line).

The comparisons discussed above were all based on the assumption that the photons are coupled to a single trapped atom. We now generalize these results to the case of $N$ atoms. The red solid curve in Fig. 8 shows the success probability for a nonlinear phase gate as a function of $N$, where the parameters $\delta_r$ and $\Delta_r$ were optimized for each value of $N$. The black dashed curve shows the success probability for a quantum Zeno gate as a function of $N$, where the values of $\varepsilon_\kappa$ and $\Delta_r$ were also optimized for each value of $N$. All of these results correspond to $\Omega = 50$ and $\Gamma_r = 0.1$ as before.

We can see from Fig. 8 that the optimized probability of success for both gates increases as the number of atoms increases. In the case of nonlinear phase gates, increasing the number of atoms reduces the required interaction time and thus reduces the effects of cavity loss. This effect saturates, however, at sufficiently large values of $N$ where losses due to atomic decay dominate. A similar situation occurs for quantum Zeno gates, where larger numbers of atoms give a larger two-photon absorption rate, which also allows faster gate operation and reduces the effects of cavity loss. It can be seen once again that the optimized performance of the two kinds of gates are very similar under the same experimental conditions. Comparing Figs. 7 and 8 at a value of $\Omega = 50$ suggests that increasing the number of atoms from 1 to 100 for a cavity with moderate loss ($\Gamma_r = 0.1$) allows similar performance to that obtained using a single atom in a much better cavity ($\Gamma_r = 1.8$).

The performance of both gates could be further improved if the atomic Rabi frequency were increased by reducing the mode volume of the resonator, for example. Optical nonlinearities are much stronger for classical fields with larger numbers of photons, and all-optical switching of classical fields based on the Zeno effect can be very efficient [39].

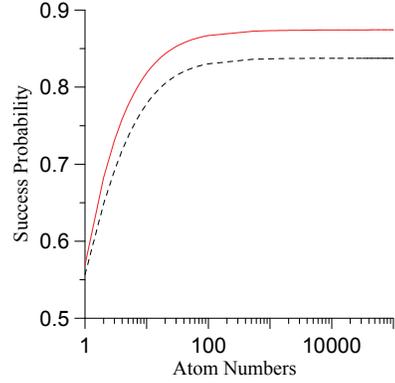

Fig. 8. Optimized success probability for nonlinear phase gates (red solid line) and quantum Zeno gates (black dashed line) as a function of the number of atoms that are coupled to the photons. These results assume the same resonator properties in both cases, namely $\Gamma_r = 0.1$ and $\Omega = 50$.

## VI. SUMMARY AND CONCLUSIONS

We have compared the expected performance of nonlinear phase gates based on the Kerr effect with that of quantum Zeno gates based on the use of strong two-photon absorption. Both of these effects rely on nonlinear effects in three-level atoms, where the upper atomic level is detuned in the case of the nonlinear phase gates while it is on resonance for the operation of quantum Zeno gates. All of the fixed hardware parameters were assumed to be the same in both cases, while any adjustable parameters were chosen to optimize the performance of each gate separately.

The optimized performance of these two kinds of gates was found to be comparable over a wide range of conditions. Perhaps this is not too surprising, since the operation of one gate depends on nonlinear changes in the real part of the index of refraction while the other depends on changes in the imaginary part. Zeno gates show a small advantage for low values of the atomic Rabi frequency and small numbers of atoms, while nonlinear phase gates give slightly better performance in the opposite limit. It is worth noting that Zeno gates do not require careful control of the magnitude of the nonlinear interaction, where the highest possible two-photon absorption rate is all that is required. In addition, Figs. 6 and 7 suggest that improved cavity performance is probably the best path forward for achieving higher fidelities for both nonlinear phase gates and Zeno gates

We would like to acknowledge valuable discussions with Todd Pittman and Scott Hendrickson. This work was supported in part by the Intelligence Advanced Research Projects Activity (IARPA) under United States Army Research Office (USARO) contract W911NF-05-1-0397, and the National Science Foundation (NSF) under grant 0652560.

**APPENDIX. SWITCHING OF SINGLE PHOTONS BETWEEN WAVEGUIDES AND RESONATORS**

It was assumed in the main text that single photons can be coupled from a waveguide into a toroidal resonator and then switched back out again after the desired quantum logic operation has been performed. The main goal of this paper is to illustrate the fundamental connection between these two types of logic devices, and the details of the coupling of the photons from the wave guide into the resonator is not our primary focus. Nevertheless, we show in this Appendix that single photons can be switched into and out a resonator if the coupling between the waveguide and the resonator can be controlled as a function of time. This could be accomplished by varying the index of refraction in the coupling region using external control fields or by changing the distance between the waveguide and the resonator using MEMS technology, for example.

The switching operation is a linear process and there is no fundamental difference between a classical and a quantum-mechanical description. For simplicity, we will consider classical field amplitudes here, although the same results apply to second-quantized field operators. We will consider the case in which the switching time is much slower than the time required for the field to propagate once around the toroidal resonator. This corresponds to the quasi-static limit in which the field intensity is essentially uniform throughout the resonator.

We will begin with the case in which a photon is initially present inside the resonator with no coupling to the waveguide. It will be shown that an appropriate choice of a time-dependent coupling allows the photon to be switched out of the resonator and into the waveguide with a Gaussian pulse shape. From time-reversal invariance, this process can be reversed to couple a photon with a Gaussian pulse shape into the resonator. It will be found that the efficiency of this process can approach 100% in the limit of a slow switching time, with relatively low losses for faster switching times.

The system of interest is shown in Fig. A1. The electric field $E_R(x,t)$ in the resonator will be written in the form $E_R(x,t) = E(t)e^{ikx}e^{-i\omega t}$. Here $x$ is the distance from the coupling region, $k$ is the wave vector, $\omega$ is the frequency, and $E(t)$ is a real function of time. The field is single-valued so that $E_R(L,t) = E_R(0,t)$, where $L$ is the circumference of the resonator. (For simplicity, we are ignoring the vector nature of the field mode and considering only its amplitude as a function of x, which does not affect the result.) The electric field in the output of the waveguide will be denoted by $E_A(y,t)$, where $y$ is the distance from the coupling region. The electic field coupled out of the toroid can be described by a real coupling coefficient $R(t)$, where $E_A(0,t) = iRE_R(0,t)$. This coupling can be viewed as being analogous to the operation of a beam splitter and our approach is somewhat similar to that of Ref. [39].

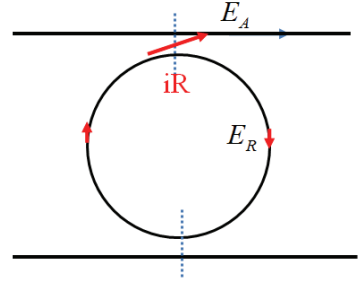

Fig.A1. Coupling of a toroidal resonator into the upper tapered fiber or waveguide with a time-dependent coupling coefficient $R(t)$. In this case, a photon is assumed to be in the resonator initially and the goal is to couple it into the waveguide in the form of a Gaussian wave packet. (There is no coupling to the lower waveguide.)

The electric field propagating in the waveguide satisfies the condition

$$E_A(y,t) = E_A(0, t - y/c) \qquad (A1)$$

Here $c$ is the velocity of the photon in the waveguide, which is assumed to be the same as that in the resonator. We require that the output be a Gaussian pulse of the form

$$E_A(0,t) = iE_1 \exp[-t^2/a^2]\exp[-i\omega t]. \qquad (A2)$$

where the constant $E_1$ is taken to be real. The initial value of the field inside the resonator is assumed to be given by

$$E_R(0,-\infty) = E_0 \qquad (A3)$$

where $E_0$ is also a real constant.

The losses in the cavity are assumed to be negligible over the switching time and we can assume that the cross-section areas of the waveguide and the toriod are the same without any loss of generality. Conservation of energy then gives

$$\frac{1}{2}\varepsilon E_0^2 L = \frac{1}{2}\varepsilon E^2(t)L + \frac{1}{2}\varepsilon \int_0^{c(t-t_0)} |E_A(y,t)|^2 \, dy \quad (A4)$$

Here $\varepsilon$ is the permittivity of the material in the waveguide and toroid.

The integral $I(t)$ in Eq. (A4) is given by

$$I(t) = \int_0^{c(t-t_0)} |E_A(y,t)|^2 \, dy = \int_0^{c(t-t_0)} |E_A(0,t-y/c)|^2 \, dy =$$
$$E_1^2 \int_0^{c(t-t_0)} Exp[-2(y/c-t)^2/a^2] \, dy. \quad (A5)$$

Making the change of variables $z = y/c - t$ gives

$$I(t) = cE_1^2 \int_{-t}^{\infty} Exp[-\frac{2z^2}{a^2}] \, dy = cE_1^2 \frac{a}{2}\sqrt{\frac{\pi}{2}}\left(1 + Erf[\frac{\sqrt{2}t}{a}]\right). \quad (A6)$$

For reasons that will become apparent, we assume that at the end of the switching process ($t \to \infty$) there is a small field amplitude $\Delta E$ still left in the resonator, which corresponds to a residual energy of $\varepsilon(\Delta E)^2 LA/2$. Here $A$ is the effective cross-sectional area of the resonator, which can take into account the actual intensity distribution of the field modes [32]. Energy conservation at the end of the switching process combined with Eq. A6 then gives

$$\frac{1}{2}\varepsilon E_0^2 L = \frac{1}{2}\varepsilon \lim_{\substack{t_0 \to -\infty \\ t \to \infty}} cE_1^2 \frac{a}{2}$$
$$\times \sqrt{\frac{\pi}{2}}\left(Erf\left[\frac{\sqrt{2}t}{a}\right] - Erf\left[\frac{\sqrt{2}t_0}{a}\right]\right) + \frac{1}{2}\varepsilon(\Delta E)^2 L \quad (A7)$$
$$= \frac{1}{2}\varepsilon\left(cE_1^2 a\sqrt{\frac{\pi}{2}} + (\Delta E)^2 L\right)$$

Thus the constants $E_0$ and $E_1$ are related by

$$E_0^2 = \sqrt{\frac{\pi}{2}}\frac{a}{\tau_R} E_1^2 + (\Delta E)^2, \quad (A8)$$

which can be solved for the value of the constant $E_1$. Here $\tau_R = L/c$ is the time required for a photon to travel once around the circumference of the resonator. Inserting Eqs. (A6) and (A8) into Eq. (A4) gives

$$E^2(t) = \sqrt{\frac{\pi}{8}}\frac{a}{\tau_R} E_1^2 \left(1 - Erf\left[\frac{\sqrt{2}t}{a}\right]\right) + \Delta E^2 \quad (A9)$$

This gives the field remaining in the resonator as a function of time.

We can now solve for the coupling coefficient $R(t)$ required to give the desired output pulse shape. From Eq. (A2) we find that

$$R(t) = E_1 Exp\left[-\frac{t^2}{a^2}\right] / E(t). \quad (A10)$$

Inserting Eq. (A9) into Eq. (A10) gives

$$R^2(t) = \frac{Exp\left[-2t^2/a^2\right]}{\sqrt{\frac{\pi}{8}}\frac{a}{\tau_R}\left(1 - Erf\left[\frac{\sqrt{2}t}{a}\right]\right) + \left(\frac{\Delta E}{E_1}\right)^2}$$
$$= \frac{Exp\left[-2t^2/a^2\right]}{\sqrt{\frac{\pi}{8}}\frac{a}{\tau_R}\left(1 - Erf\left[\frac{\sqrt{2}t}{a}\right]\right) + \left(\frac{E_0}{E_1}\right)^2\left(\frac{\Delta E}{E_0}\right)^2} \quad (A11)$$

From Eq. (A8) it follows that

$$\left(\frac{E_0}{E_1}\right)^2 = \sqrt{\frac{\pi}{2}}\frac{a}{\tau_R}\frac{1}{1-r} \quad (A12)$$

where $r = (\Delta E/E_0)^2$ is the ratio of the energy left in the resonator to the initial energy. Inserting Eq. (A12) into Eq. (A11) finally gives

$$R^2(t) = \frac{Exp\left[-2t^2/a^2\right]}{\sqrt{\frac{\pi}{2}}\frac{a}{\tau_R}\left(\frac{r}{1-r} + \frac{1}{2} - \frac{1}{2}Erf\left[\frac{\sqrt{2}t}{a}\right]\right)} \quad (A13)$$

as the required coupling coefficient.

Fig. A2a shows a plot of the square of the coupling coefficient as a function of time for several different values of the residual energy parameter $r$. All of these results correspond to an output pulse width corresponding to $a = 20\sqrt{2}\tau_R$. It can be seen that the necessary value of the coupling coefficient can be reduced by leaving a larger fraction of the energy in the resonator, if desired. A relatively small loss of $r = 0.01\%$ allows a maximum value of $R^2 = 0.22$, for example. Even smaller values of $R$ can be achieved for longer switching times. Fig. A2b shows a plot of the corresponding electric field amplitude

in the output waveguide as a function of time for comparison

It follows from time-reversal invariance that a Gaussian pulse can be switched into the resonator if the time-dependence of the coupling coefficient is reversed:

$$R(t) \rightarrow R(-t) . \quad (A14)$$

These results show that it is possible in principle to switch a photon into or out of a resonator as assumed in the text, provided that the coupling coefficient can be controlled as a function of time.

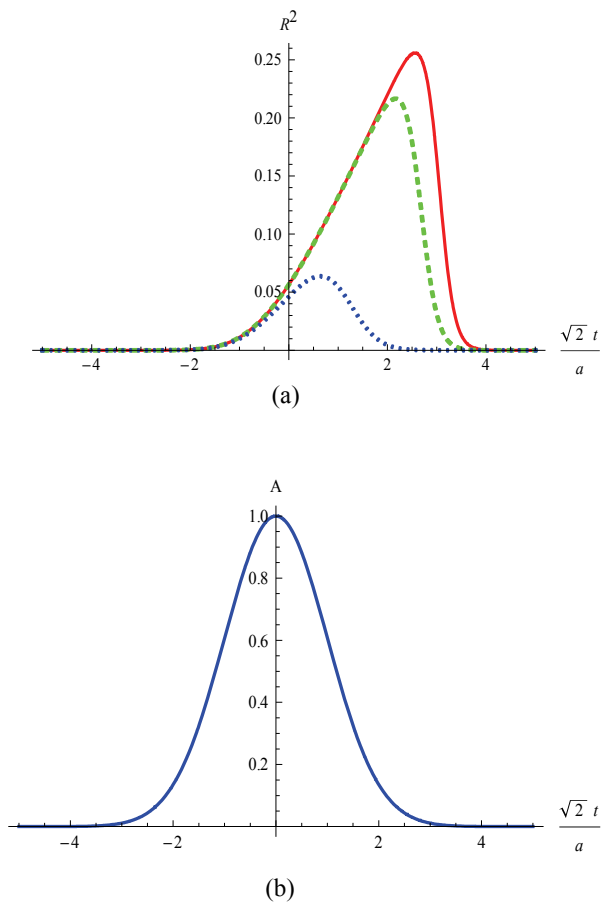

Fig. A2. Typical results for the required coupling coefficient $R(t)$. (a) The square of the required coupling coefficient as a function of time for a pulse width corresponding to $a = 20\sqrt{2}\tau_R$. The red (solid), green (dashed), and blue (dotted) curves correspond to $r = 0.001\%$, $0.01\%$, and $10\%$, respectively. (b) The amplitude of the electric field in the output waveguide as a function of time for comparison. The time-dependence of the coupling coefficient was chosen to produce a Gaussian profile.

**REFERENCES**

1. J. I. Cirac and P. Zoller, Phys. Rev. Lett. 74, 4091 (1995); C. Monroe, D.M. Meekhof, B.E. King, W.M. Itano, and D.J. Wineland, ibid. 75, 4714 (1995); T. Monz, K. Kim, A. S. Villar, P. Schindler, M. Chwalla, M. Riebe, C. F. Roos, H. Häffner, W. Hänsel, M. Hennrich, and R. Blatt, ibid. 103, 200503 (2009).
2. D. Jaksch, H.-J. Briegel, J. I. Cirac, C. W. Gardiner, and P. Zoller, Phys. Rev. Lett. 82, 1975 (1999); L. Isenhower, E. Urban, X. L. Zhang, A. T. Gill, T. Henage, T. A. Johnson, T. G. Walker, and M. Saffman, Phys. Rev. Lett. 104, 010503 (2010).
3. D. DeMille, Phys. Rev. Lett. 88, 067901 (2002); A. André, D. DeMille, J. M. Doyle, M. D. Lukin, S. E. Maxwell, P. Rabl, R. J. Schoelkopf and P. Zoller, Nature Physics 2, 636 (2006); P. Rabl, D. DeMille, J. M. Doyle, M. D. Lukin, R. J. Schoelkopf and P. Zoller, Phys. Rev. Lett. 97, 033003 (2006); S. F. Yelin, K. Kirby, and R. Côté, Phys. Rev. A 74, 050301(R) (2006).
4. T. Yamamoto, Yu. A. Pashkin, O. Astafiev, Y. Nakamura and J. S. Tsai, Nature 425, 941 (2003); J. H. Plantenberg, P. C. de Groot, C. J. P. M. Harmans & J. E. Mooij, ibid. 447, 836 (2007); A. O. Niskanen, K. Harrabi, F. Yoshihara, Y. Nakamura, S. Lloyd, J. S. Tsai, Science 316, 723 (2007); Erik Lucero, M. Hofheinz, M. Ansmann, Radoslaw C. Bialczak, N. Katz, Matthew Neeley, A. D. O'Connell, H. Wang, A. N. Cleland, and John M. Martinis, Phys. Rev. Lett. 100, 247001 (2008); L. DiCarlo, J. M. Chow, J. M. Gambetta, Lev S. Bishop, B. R. Johnson, D. I. Schuster, J. Majer, A. Blais, L. Frunzio, S. M. Girvin and R. J. Schoelkopf, Nature 460, 240 (2009).
5. D. Loss and D.P. DiVincenzo, Phys. Rev. A 57, 120 (1998); B. E. Kane, Nature 393, 133 (1998); K. J. Xu, Y. P. Huang, M. G. Moore, and C. Piermarocchi, Phys. Rev. Lett. 103, 037401 (2009).
6. G. J. Milburn, Phys. Rev. Lett. 62, 2124 (1989).
7. Q. A. Turchette, C. J. Hood, W. Lange, H. Mabuchi, and H. J. Kimble, Phys. Rev. Lett, 75, 4710 (1995).
8. E. Knill, R. Laflamme, and G. J. Milburn, Nature 409, 46 (2001).
9. T. B. Pittman, B. C. Jacobs, and J. D. Franson, Phys. Rev. A 64, 062311 (2001).
10. T. B. Pittman, B. C. Jacobs, and J. D. Franson, Phys. Rev. Lett 88, 257902 (2002).
11. J. D. Franson, M. M. Donegan, M. J. Fitch, B. C. Jacobs, and T. B. Pittman, Phys. Rev. Lett 89, 137901 (2002).
12. T. B. Pittman, M. J. Fitch, B. C. Jacobs, and J. D. Franson, Phys. Rev. A 68, 032316 (2003).
13. J. L. O'Brein, G. J. Pryde, A. G. White, T. C. Ralph, and D. Branning, Nature 426, 264 (2003).
14. R. Prevedal, P. Walther, F. Tiefenbacher, P. Bohi, R. Kaltenbaek, T. Jennewein, and A. Zeilinger, Nature 445, 65 (2007).

15. K. Nemoto and W. J. Munro, Phys. Rev. Lett 93, 250502 (2004).
16. T. P. Spiller, K. Nemoto, S. L. Braunstein, W. J. Munro, P. Van Loock, and G. J. Milburn, New Journal of Physics 8, 30 (2006).
17. K. M. Birnbaum, A. Boca, R. Miller, A. D. Boozer, T. E. Northup and H. J. Kimble, Nature 436, 87 (2005).
18. S. Lloyd and S. L. Braunstein, Phys. Rev. Lett. 82, 1784 (1999).
19. Nicolas C. Menicucci, Peter van Loock, Mile Gu, Christian Weedbrook, Timothy C. Ralph, and Michael A. Nielsen, Phys. Rev. Lett. 97, 110501 (2006).
20. Mile Gu, Christian Weedbrook, Nicolas C. Menicucci, Timothy C. Ralph, and Peter van Loock, Phys. Rev. A 79, 062318 (2009).
21. J. D. Franson, B. C. Jacobs, and T. B. Pittman, Phys. Rev. A 70, 062302 (2004).
22. J. D. Franson, B. C. Jacobs, and T. B. Pittman, J. Opt. Soc. Am. B 24, 209 (2007).
23. P. M. Leung and T. C. Ralph, Phys. Rev. A 74, 062325 (2006).
24. Patrick M Leung and Timothy C Ralph, New J. Phys. 9, 224 (2007).
25. Casey R. Myers and Alexei Gilchrist, Phys. Rev. A 75, 052339 (2007).
26. Y. P. Huang and M. G. Moore, Phys. Rev. A 77, 062332 (2008).
27. Hao You, S. M. Hendrickson, and J. D. Franson, Phys. Rev. A 80, 043823 (2009).
28. D. K. Armani, T. J. Kippenberg, S. M. Spillane and K. J. Vahala, Nature 421, 925 (2003).
29. K. J. Vahala, Nature 424, 839 (2003).
30. T. J. Kippenberg, S. M. Spillane, D. K. Armani, and K. J. Vahala, Applied Physics Letters 83(4), 797 (2003).
31. S. M. Spillane, T. J. Kippenberg, K. J. Vahala, K. W. Goh, E. Wilcut, and H. J. Kimble, Phys. Rev. A 71, 013817 (2005).
32. Bumki Min, Lan Yang, and Kerry Vahala, Phys. Rev. A 76, 013823 (2007).
33. Hao You, S. M. Hendrickson, and J. D. Franson, Phys. Rev. A 78, 053803 (2008).
34. C. Cohen-Tannoudji, "Optical pumping and interactions of atoms with the electromagnetic field", in "Cargese Lectures in Physics", Vol 2, pp.347-393, ed. by M. Levy (Gordon and Breach, 1968).
35. C. Cohen-Tannoudji, J. Dupont-Roc, G. Grynberg, in "Atom-Photon Interactions : Basic Processes and Applications" (Wiley, New-York, 1992).
36. N. Bloembergen and M. D. Levenson, "Doppler-free two-photon absorption spectroscopy" , in "Topics in Applied Physics", Vol 13, pp.329, ed. by K. Shimoda (Springer Verlag, 1976).
37. H. J. Kimble, Physica Scripta T76, 127 (1998).
38. M. A. Nielsen and I. L. Chuang, Quantum Computation and Quantum Information (Cambridge University Press, Cambridge, 2000).
39. B.C. Jacobs and J.D. Franson, Phys. Rev. A **79**, 063830 (2009).